\documentclass[fleqn,usenatbib]{mnras}

\usepackage{newtxtext,newtxmath}

\usepackage[T1]{fontenc}

\DeclareRobustCommand{\VAN}[3]{#2}
\let\VANthebibliography\thebibliography
\def\thebibliography{\DeclareRobustCommand{\VAN}[3]{##3}\VANthebibliography}

\usepackage{graphicx}	
\usepackage{amsmath}	
\usepackage{amssymb}	

\title[SB2s in M~11]{Double-lined spectroscopic binaries in M11.}

\author[M. Kovalev et al.]{
Mikhail Kovalev$^{1,2}$\thanks{E-mail: mikhail.kovalev@ynao.ac.cn},
Ilya Straumit$^{3,4}$
\\
$^{1}$Yunnan Astronomical Observatories, China Academy of Sciences, 650216 Kunming, China\\
$^{2}$Max Planck Institute for Astronomy, D-69117 Heidelberg, Germany\\
$^{3}$Department of Astronomy, The Ohio State University, Columbus, OH 43210, USA\\
$^{4}$Institute for Astronomy, KU Leuven, Celestijnenlaan 200D bus 2401, 3001 Leuven, Belgium
}

\date{Accepted XXX. Received YYY; in original form ZZZ}

\pubyear{2021}

\def\kms{\,{\rm km}\,{\rm s}^{-1}}

\def\feh{\hbox{[Fe/H]}}

\newcommand{\teff}{T_{\rm eff}}
\newcommand{\rv}{{\rm RV}}
\def\Vmic{V_{\rm mic}}

\def\vsini{v{\rm \sin i}}
\def\logg{\log({\rm g})}
\def\snr{\hbox{S/N}}

\begin{document}
\label{firstpage}
\pagerange{\pageref{firstpage}--\pageref{lastpage}}
\maketitle

\begin{abstract}
We have developed a new method for spectral analysis of binaries.
Our method successfully identifies SB2 candidates from high-resolution Gaia-ESO spectra. Compared to the commonly used cross-correlation function analysis, it works for binaries with rapidly rotating components. We test our method on synthetic and observational spectra of BAFG-stars with $\vsini$  from 1 to $330~\kms$ in the open cluster M~11. We confirm five previously detected SB2 candidates and find 19 new ones. For three SB2 candidates we find circular orbits and obtain dynamical mass ratios.
\end{abstract}




\section{Introduction}

M~11 (NGC~6705), also known as the Wild Duck Cluster, is an open cluster from the constellation Scutum. 
 \citet{recentgaia} provides the most recent estimates of the age $294\pm 1.2$ Myr, the distance $1889\pm65$ pc and the extinction $A_v=1.457\pm0.051$ mag based on Gaia DR2 data \citep{gdr2}. 
This cluster is used for the internal calibration of the Gaia-ESO spectroscopic survey \citep{Pancino2017}, so M~11 stars have many high-resolution observations. For example, \citet{cantat2014} used Gaia-ESO spectra to determine the metallicity of the cluster $\feh=0.10\pm0.06$ dex and age from 250 to 316 million years (depending on the isochrone model used). They report an average radial velocity of the cluster $\langle\rv\rangle=35.9\pm2.8\kms$ based on GIRAFFE HR15 \citep{Pasquini2002} spectra.
\citet{marino2018} found many fast rotating stars in M~11. 
In \citet{merle2017,merle2020} the authors have used Gaia-ESO spectra and cross-correlation function analysis to find many single-lined (SB1) and double-lined (SB2) spectroscopic binaries in M~11.
\par
In this paper, we use Gaia-ESO spectra of M~11 stars to test our method of analysing the composite spectra of binary stars. Our method is inspired by the seminal \citet{bardy2018} work, where the authors analysed infrared spectra using a composite spectral model, based on data-driven spectral model and grid of isochrones to scale binary components contribution. In Section~\ref{Observations} we describe the observations and methods. Section~\ref{results} presents and discusses our results. In Section ~\ref{concl} we summarise the paper and draw conclusions.

\section{Observations \& Methods}
\label{Observations}
\subsection{Observations}

We have used the spectra of 265 M~11 stars observed as part of the Gaia-ESO spectroscopic survey \citep{Gilmore2012, Randich2013}. These spectra are now publicly available as part of the third data release (DR3.1)\footnote{\url{http://archive.eso.org/wdb/wdb/adp/phase3_spectral/form?collection_name=GAIAESO}}. The data were obtained with the GIRAFFE \citep{Pasquini2002} instrument on the VLT (Very Large Telescope) of the European Southern Observatory (ESO). We used spectra taken with the HR21 setup, which covers 505 \AA~ from 8475 \AA~ to 8980 \AA~, at a resolution of $R=\lambda/\Delta \lambda \sim 16\,200$. This setting is used because of the strong Ca~II lines visible in hot stars. The average signal-to-noise ratio ($\snr$) of the spectrum ranges from 30 to 120 ${\rm pix}^{-1}$, with most spectra having $\snr$ values in the 40-60 ${\rm pix}^{-1}$ range. All spectra contain only one exposure taken on one of the two nights MJD=56103.199~day (158 spectra) and MJD=56442.366~day (107 spectra). This dataset includes six SB2 candidates from \citet{merle2017}, 10 SB1 candidates from \citet{merle2020}, 
 one detached eclipsing binary (EB) KV~29 \citep{kv29} and five other variables, based on SIMBAD database. We found no matches with SB9 catalogue of spectroscopic binary orbits \citep{sb9}.


\subsection{Spectral models}
The synthetic spectra for single stars were generated using GSSP (Grid Search in Stellar Parameters) software \citep{2015A&A...581A.129T}. GSSP uses SynthV code for spectral synthesis \citep{1996ASPC..108..198T}, which is based on LTE (local thermodynamic equilibrium) approach and uses a list of spectral lines from VALD database \citep{2015PhyS...90e4005R}. Stellar atmosphere models that were used to produce synthetic spectra were precomputed using the LL code \citep{2004A&A...428..993S}. The distinctive feature of the LL code is the use of "line-by-line" (LL) technique to compute the line absorption coefficients taking into account opacities caused by nearby lines at both sides from a given wavelength point. This method for computing line absorption was not used regularly due to high computational demands. Previous attempts to introduce line-by-line approach were realized, for example, by \citet{1965ApJ...141...73M}, but only for a limited number of lines and only in the region covered by Lyman series. In contrast, LL models are computed based on a full line list from the VALD database, which is made possible in part by optimized numerical techniques implemented in LL code, and in part by the increased performance of modern computers. This makes the LL method free from approximations inherent to traditional techniques such as opacity distribution function method \citep{1979ApJS...40....1K} or opacity sampling method \citep{1986A&A...167..304E, 1989ApJ...345.1014A}. In comparison, MARCS model grid \citep{2008A&A...486..951G} is computed using opacity sampling method and contains atmospheres for $\mathrm{T_{eff} \leq 8000~K}$ (LL model grid extends to $\mathrm{T_{eff} \leq 34000~K}$); ATLAS9 model grid \citep{2003IAUS..210P.A20C} is computed using opacity distribution function method and extends to $\mathrm{T_{eff} \leq 50000~K}$ (depending on $\logg$).
\par
The grid of models (70686 in total) was computed for a range of  $\teff$ between 5000 and 15000~K in steps of 500~K, $\logg$ between 3.0 and 5.0 in steps of 0.2 (cgs units),  $\vsini=1\,\kms$ plus $\vsini$ from 10 to 330 $\kms$ in steps of 20 $\kms$ and [Fe/H] between $-$0.8 and $+$0.8 dex with steps of 0.1 dex. Microturbulence was fixed to $\Vmic=2~\kms$. All models have been degraded to resolution $R=16\,200$ and resampled to a regular wavelength scale with a step of $0.05$ \AA~. The single-star spectrum was generated using simple linear interpolation within the grid. 

\subsection{Method}
\label{sec:maths} 

The normalised binary spectrum is generated as the sum of two normalised single-star spectra scaled according to the difference in luminosity, which is a function of $\teff$ and stellar size. We use the following equation:    

\begin{align}
    &{f}_{\lambda,{\rm binary}}=\frac{{f}_{\lambda,1} + k_\lambda {f}_{\lambda,2}}{1+k_\lambda}.\\
    &k_\lambda= \frac{B_\lambda(\teff{_{1}})~m_1}{B_\lambda(\teff{_{2}})~m_2} 10^{\logg_{2}-\logg_{1}}
	\label{eq:bolzmann}
\end{align}
 where  $k_\lambda$ is the luminosity ratio per wavelength unit, $B_\lambda$ is the black body radiation, $\teff{_{i}}$ is the effective temperature, $\logg_i$ is the surface gravity, $m_i$ is the mass and ${f}_{\lambda,i}$ is the flux, Doppler-shifted with $\rv_i$ for component $i$. 
 If both components have all the same parameters, the binary model is identical to the single-star model.
 \par
 Throughout the paper we always assume primary component as the brighter one. The mass ratio is defined as $q=\frac{m_1}{m_2}$, therefore if primary component is heavier we have $q > 1$. Thus our mass ratio is inverted in comparison with a traditional definition of the mass ratio $Q=1/q$, which is common in the literature. 
\par
The model spectrum is later multiplied by a normalisation function, which is a linear combination of the first four Chebyshev polynomials, defined on wavelength interval $\lambda$ \citep[similar to][]{kovalev19}. The resulting spectrum is compared with the observed one, previously divided by its median value, using least squares fitting. The normalisation function is updated simultaneously with the other fitting parameters. Compared to traditional normalisation methods, this method has minimal human intervention: only the appropriate Chebyshev polynomial order needs to be set (in our case four). The disadvantages of this method are the increased computational time and resources required for optimisation. It will also only work if the model spectrum can adequately represent all features of the observed spectrum, and the response between observation and model can be approximated by a smooth, continuous function.  We use the \texttt{Python} function \texttt{scipy.optimise.curve\_fit}, which provides the optimal spectral parameters and radial velocities of each component, the mass ratio and four coefficients of the Chebyshev polynomials. We keep the metallicity equal for both components in the binary system. We have a total of $p=14$ free parameters for the binary model fit and $p=9$ free parameters for the single-star model fit. We estimate goodness of the fit parameter by reduced $\chi^2$:

\begin{flalign}
\label{eq:chi2}
   \chi^2 =\frac{1}{N-p} \sum \left[ \left({f}_{\lambda,{\rm observed}}-{f}_{\lambda,{\rm model}}\right)/{\sigma}_{\lambda}\right]^2
\end{flalign}
where $N$ is a number of wavelength points in the observed spectrum. 

\par
 At first we analyse observed spectrum with the single-star model with four random optimiser initialisations, to explore the parameter space and avoid local minima. The solution with minimal $\chi^2$ is chosen as a single-star result. Then we run the optimisation using the binary model with seven different optimiser initialisations, changing the mass ratio and radial velocities for the components: $\Delta\rv=0,30,60,90\,\kms$ around $\rv$ from single-star fit with $q=1.01$ and $q=1.2,1.5,2.0$ with $\Delta\rv=0\,\kms$. All remaining parameters are initialised using single-star solution. As a final binary result, we choose the solution with the minimal $\chi^2$.
\par
Using these two solutions, we compute an improvement factor using Equation~\ref{eq:f_imp}, similar to \citet{bardy2018}. This improvement factor estimates the absolute difference between the two fits, and weights it by the difference between the two solutions.

\begin{align}
\label{eq:f_imp}
f_{{\rm imp}}=\frac{\sum\left[ \left(\left|{f}_{\lambda,{\rm single}}-{f}_{\lambda}\right|-\left|{f}_{\lambda,{\rm binary}}-{f}_{\lambda}\right|\right)/{\sigma}_{\lambda}\right] }{\sum\left[ \left|{f}_{\lambda,{\rm single}}-{f}_{\lambda,{\rm binary}}\right|/{\sigma}_{\lambda}\right] },
\end{align}
where ${f}_{\lambda}$ and ${\sigma}_{\lambda}$ are the observed flux and corresponding uncertainty, ${f}_{\lambda,{\rm single}}$ and ${f}_{\lambda,{\rm binary}}$ are the best-fit single-star and binary model spectra, and the sum is over all wavelength pixels. 

\subsection{Test on a simulated cluster and selection of the binary candidates}
\label{classify}
 
We have computed synthetic spectra for M~11 stars in two datasets: one with 480 simulated single stars and the other with 1480 simulated binaries. These simulated spectra are generated with parameters within the synthetic spectra grid at points randomly chosen from the PARSEC isochrone \citep{parsec1,parsec2} calculated using the cluster age  $250$ Myr and $\feh=0.10$ dex from \cite{cantat2014}. The radial velocity of the primary star is calculated using a uniform distribution in the range $\pm80\kms$  around the cluster velocity $35~\kms$. The radial velocity of the secondary component can be calculated using :
\begin{align}
    \label{eq:asgn}
    {\rm RV_2}=\gamma (1+q) - q {\rm RV_1},
\end{align}

 where $\gamma=35~\kms$ is the systemic velocity. We set $\gamma$ equal to the cluster velocity.
The models are degraded by Gaussian noise according to $\snr=50$ pix$^{-1}$. The projected rotational velocities for both components in binaries are random with a uniform distribution from 1 to $165~\kms$  if $\teff>6000$ K and from 1 to $30~\kms$ if $\teff<6000$ K. The limit for hot stars is just one half of maximal $\vsini$ in the grid. For single stars $\vsini$ are also randomly chosen in range from 1 to $30~\kms$ if $\teff<6000$ K and from 1 to $250~\kms$ if $\teff>6000$ K.  We find that Ca~II lines hardly visible in hot stars rotating faster than this limit. In the binary set $480$ spectra are binaries with identical stellar components, which will have spectra identical to the spectra of single stars in the absence of a difference in the radial velocity of the components. 
\par
\begin{figure*}
	\includegraphics[width=\textwidth]{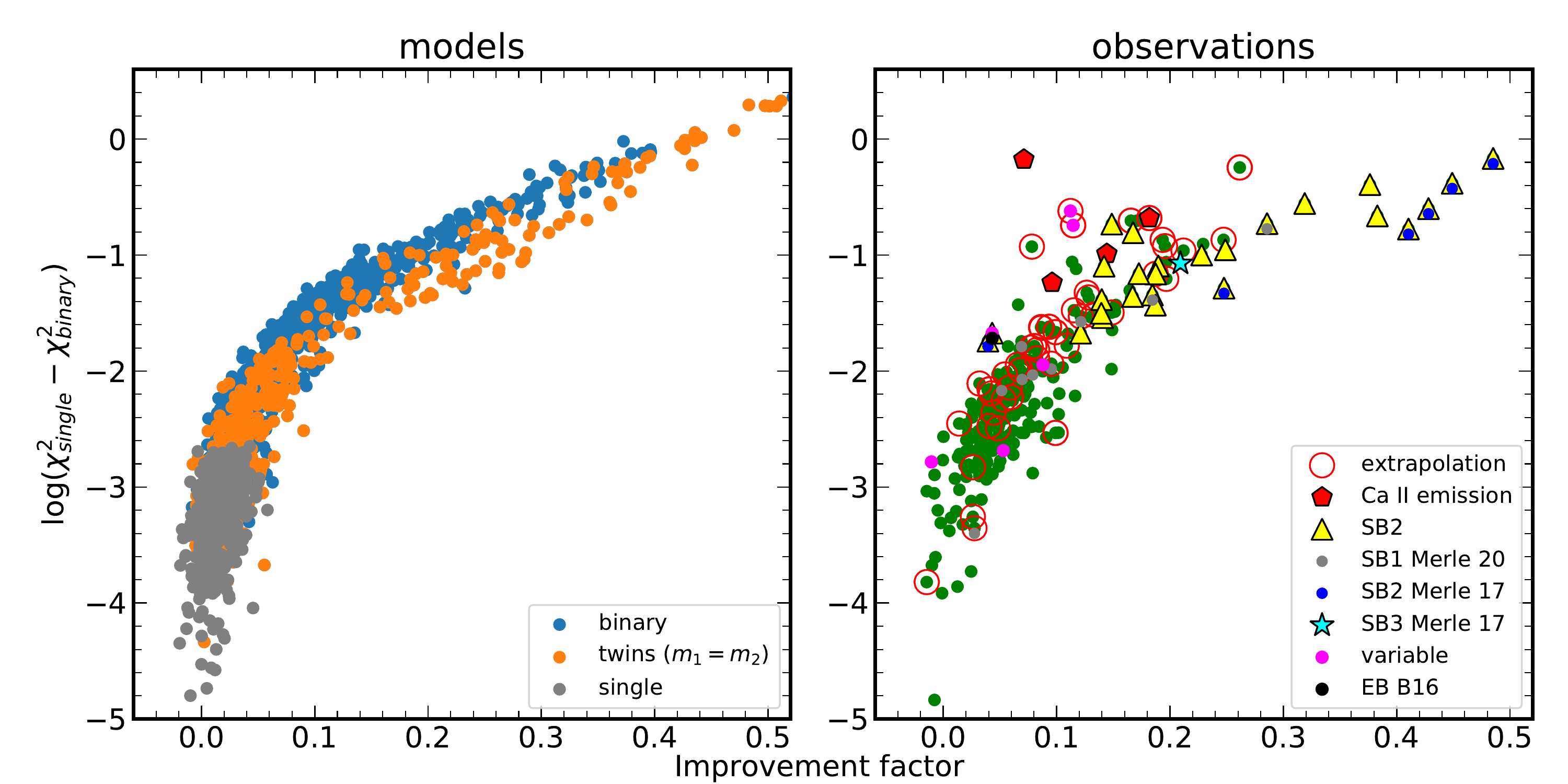}
    \caption{Selection of the binary candidates using empirical criteria based on mock stars: left panel - test on synthetic models, right panel - test on real spectra. The test on mock stars includes the analysis of single stars (gray circles), binary stars (blue circles) and twin stars (orange circles). Known variable stars from the SIMBAD database are shown as a pink and black (eclipsing binary B16 \protect\citep{kv29}) circles. Spectral binaries listed in \protect\citet{merle2017,merle2020} are shown as blue (SB2) and gray (SB1) circles. The four stars with emission in Ca~II lines are shown as red pentagons. All identified SB2 candidates are confirmed by visual inspection of the plots and are shown by yellow triangles, while all other stars are shown by green circles}
    \label{fig:sel}
\end{figure*}

We perform exactly the same analysis as for the observations on the two simulated datasets. To separate binary stars from single-star solutions, we have developed selection criteria similar to \citet{bardy2018}. The first obvious selection of single stars occurs if $\chi^2_{\rm single} < \chi^2_{\rm binary}$. This criterion selects only 2 stars from the single star dataset and zero systems from the binary ones. These two stars are fast rotators ($\vsini=135,\,230\kms$) with $\teff\sim6500$ K. For all remaining mock stars $\chi^2_{\rm single} - \chi^2_{\rm binary}>0$. 
We calculate the logarithm of this difference $\log{\Delta \chi^2}$ and plot against the  improvement factor, see the left panel of the Figure~\ref{fig:sel}. All single star layouts have $\log{\Delta \chi^2}<-2.5$, which means that this threshold can be used to select single stars reliably. However, this would only select $896$ of binary stars out of $1480$. Thus, using these criteria would allow all single stars and $\sim60\%$ of binaries to be selected reliably. Twin binaries with identical components are an interesting special case. Obviously, such stars could have spectra very similar to those of single stars and our criterion would be less reliable. We have depicted such stars with orange circles. They typically have smaller $\log{\Delta \chi^2}$ than other binary stars. Among twin binary stars, only $219$ out of $480$ have been selected as binary, and $\sim 54\%$ have not been selected. Thus, the success rate ($SR$) of our method is  ranging from $SR=\frac{219}{480}\sim0.46$ for twin binaries to $SR=\frac{896-219}{1000}\sim0.68$ for binaries with different components.
\par
\begin{figure*}
	\includegraphics[width=\textwidth]{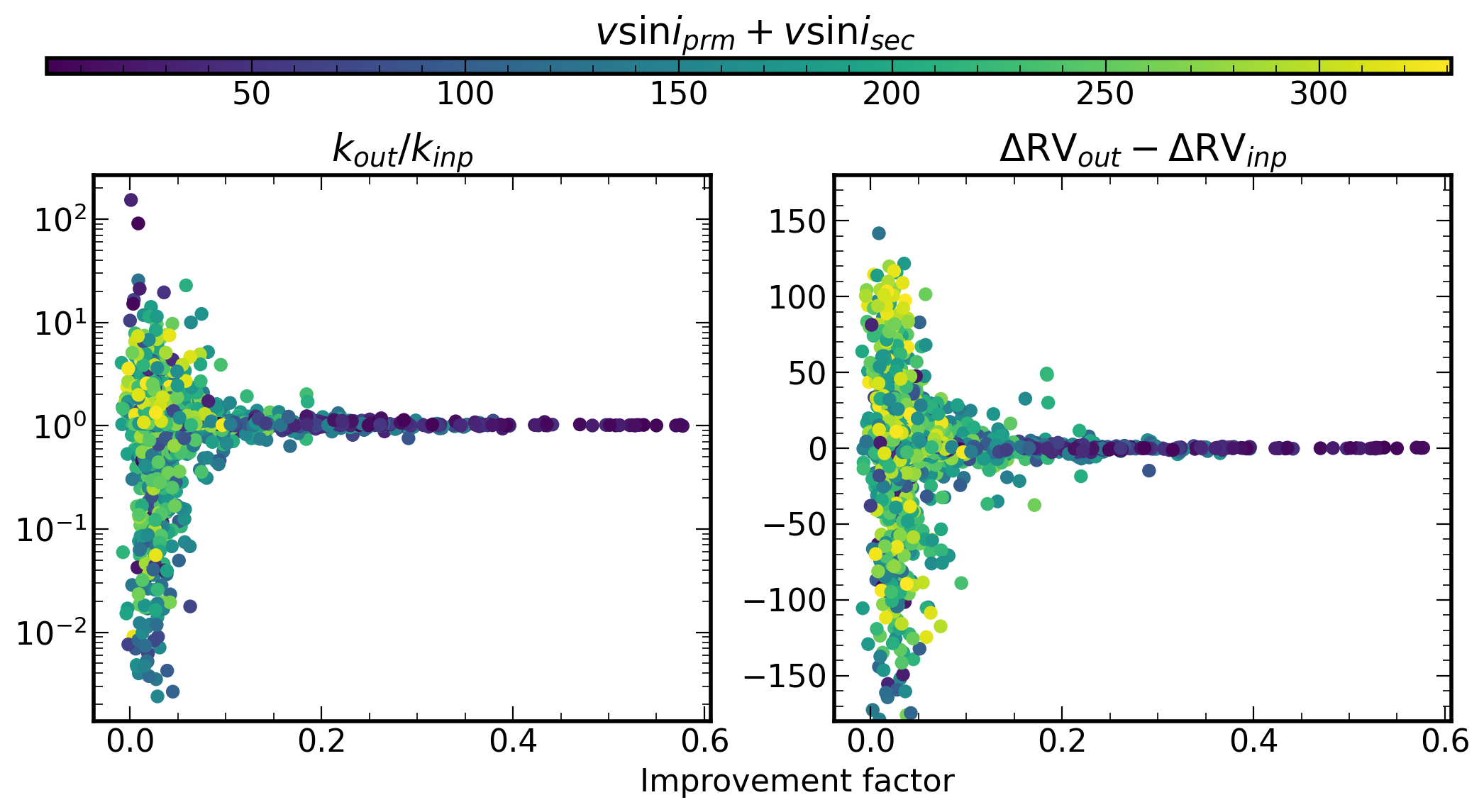}
    \caption{Recovery results for the average luminosity ratio ($k$ see Formula~\ref{eq:bolzmann}) (left panel) and radial velocity separations $\Delta {\rm RV}$ (right panel) versus $f_{\rm imp}$ for mock stars. The colour is the sum of the rotational velocities.}
    \label{fig:selimp}
\end{figure*}

We check fits for synthetic binaries and find that in many systems the recovery is wrong, for example, if the system has two similar, rapidly rotating stars or if the secondary component is too weak. Fortunately, $f_{\rm imp}$ allows us to choose good solutions. The most important thing in constructing a binary spectrum is to find the ratio of luminosities (we use the mean $k_\lambda$, see Formula~\ref{eq:bolzmann}) and the radial velocity separation ($\Delta\rv$) between the two components. In Figure~\ref{fig:selimp} we have depicted the recovery results for these quantities as a function of $f_{\rm imp}$. It can be seen that for 375 solutions with $f_{\rm imp}>0.1$ the recovery is relatively good. Binaries with slowly rotating components can be fitted much better than fast rotating ones. The maximum sum of rotational velocities of the components among the good solutions is $282\kms$.
The lowest $\Delta\rv=0.02\kms$ among good solutions is observed for a system with very different spectral components (hot rapidly rotating primary and cold secondary with narrow lines). The maximum luminosity ratio among good solutions is 22 (binary systems with primary component contribution $\sim 95\%$ (magnitude difference 3.3 mag). 

\par
\begin{figure*}
	\includegraphics[width=\textwidth]{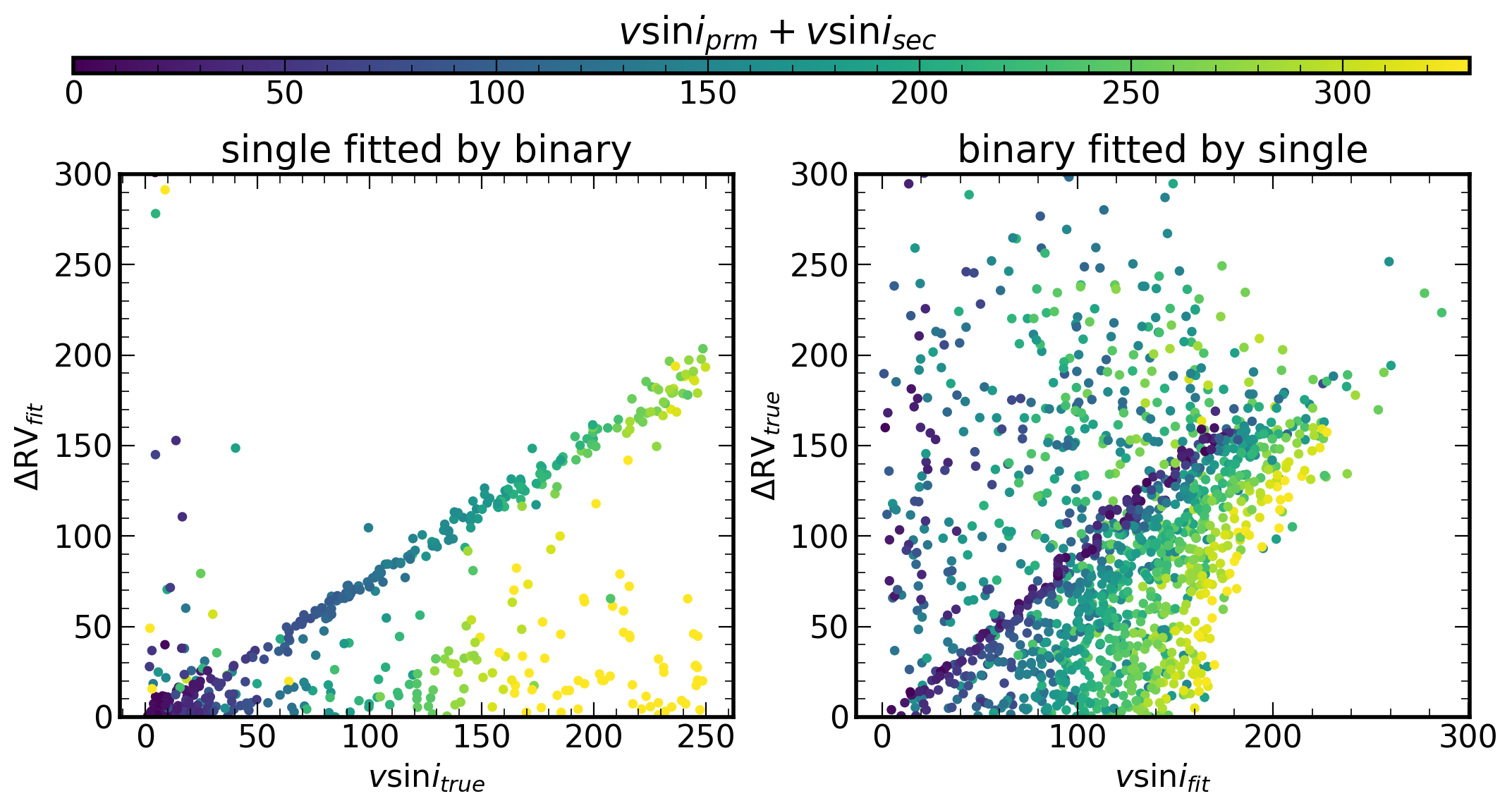}
    \caption{Degeneracy between radial velocity separations $\Delta {\rm RV}$ and $\vsini$: when single star spectrum fitted by the binary model (left panel) and when binary spectrum is fitted by the single star model (right panel). The colour is the sum of the rotational velocities in binary.}
    \label{fig:degeracy}
\end{figure*}

There is a strong degeneracy between $\Delta\rv$ and rotational velocity, see left panel in Figure~\ref{fig:degeracy}, you can see clear correlation when a rapidly rotating single star can be fitted to a binary star model consisting of two slowly rotating stars. In this case the sum $\vsini_1+\vsini_2$ is slightly larger than true $\vsini$ of the single star. The opposite effect can also occur if the single star fits well into a binary model with a rapidly rotating primary star and a much weaker secondary star, see region with small $\Delta\rv$ and high $\vsini$. Fortunately all such binary solutions have $f_{\rm imp}<0.05$ and can be easily filtered out. Another interesting result is shown on the right panel of the Figure~\ref{fig:degeracy}. We can see clear correlation between $\Delta\rv$ of binary and rotational velocity fitted by the single-star model, if the sum $\vsini_1+\vsini_2<50\,\kms$ in twin stars. This correlation can be used to identify SB2 candidates in large spectroscopic surveys if the estimated $\vsini$ is changing with time (Kovalev et al. in prep).
\par
\begin{figure*}
	\includegraphics[width=\textwidth]{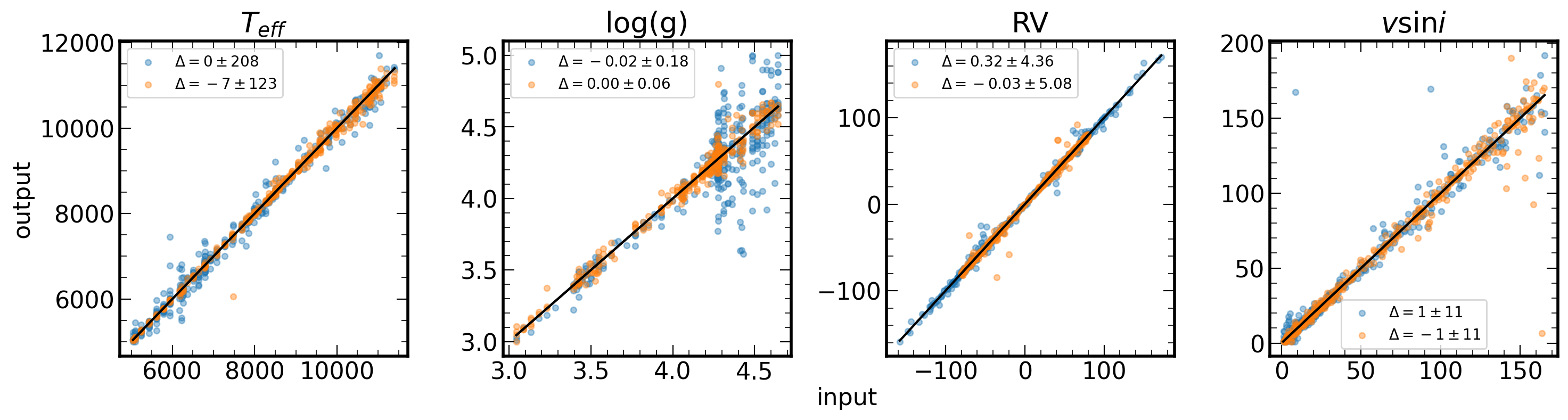}
    \caption{Recovery of the spectral parameters for the primary (orange circles) and secondary (blue circles) components in the solutions with $f_{\rm imp}>0.1$. The mean value and standard deviation of the error are shown for each parameter.}
    \label{fig:recovery}
\end{figure*}

 
We have checked how well the spectral parameters of the primary and secondary components can be recovered if the solution has $f_{\rm imp}>0.1$. Figure~\ref{fig:recovery} shows the recovery results for $\teff,\logg,\vsini$ and RV. For the primary component we have $\Delta\teff=-7\pm123$~K, $\Delta\logg=0.00\pm0.06$ cgs units, $\Delta\vsini=-1\pm11\kms$ and $\Delta {\rm RV}=-0.03\pm5.08\kms$. For the secondary component we have $\Delta\teff=0\pm207$~K, $\Delta\logg=-0.02\pm0.18$ cgs units, $\Delta\vsini=-1\pm11\kms$ and $\Delta {\rm RV}=0.32\pm4.36\kms$. The metallicity recovery $\Delta\feh=0.00\pm0.04$ dex is not shown in the Figure~\ref{fig:recovery}, as the metallicity was the same in all source stars. It is clear that the surface gravity of the secondary components is poorly recovered compared to the primary components, which may lead to incorrect estimates of the mass ratio, see Formula~\ref{eq:bolzmann}.  Therefore our results are unreliable for the mass ratio and $\logg$ of the secondary component. Additional preliminary information for the mass ratio would be very useful to constrain these parameters. Generally, radial velocities of the rapidly rotating component are less accurate than radial velocities of the slowly rotating component. 
\par 
Based on this simulation our method is able to identify SB2 candidates and derive radial velocities, $\teff,~\vsini,~\feh$ for both components and $\logg$ only for the primary. Even for $\Delta {\rm RV}=0$ we can get a good solution if the spectral components are significantly different.         

\begin{figure}%
	\includegraphics[width=\columnwidth]{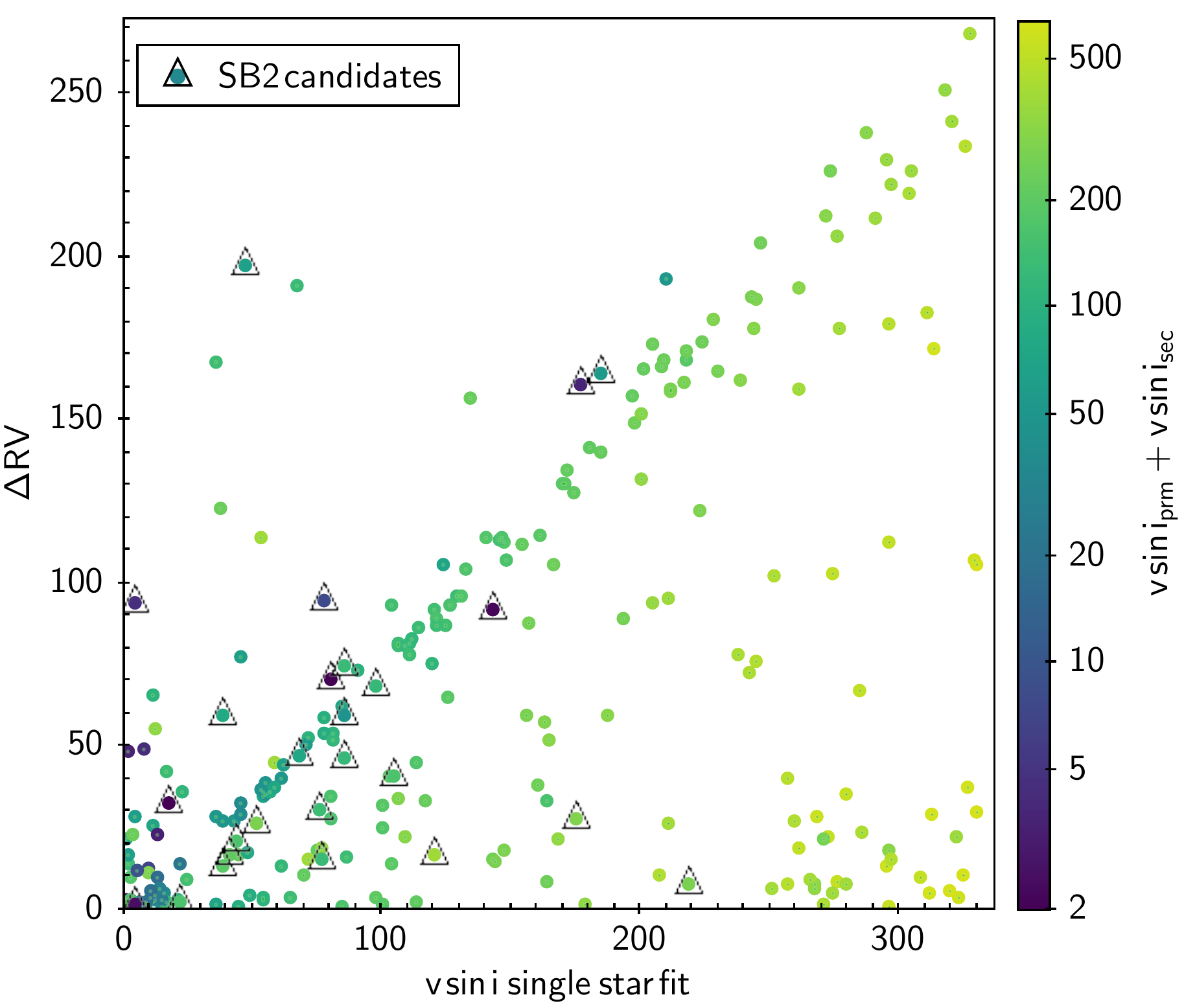}
    \caption{Same as Figure~\ref{fig:degeracy} but for observed spectra. Selected SB2 candidates are shown with open triangles.}
    \label{fig:degenobs}
\end{figure}

\begin{figure*}
	\includegraphics[width=\textwidth]{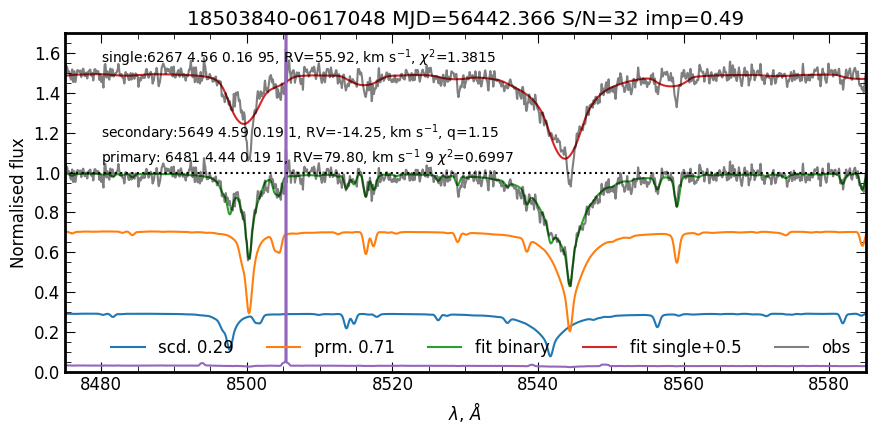}
	\includegraphics[width=\textwidth]{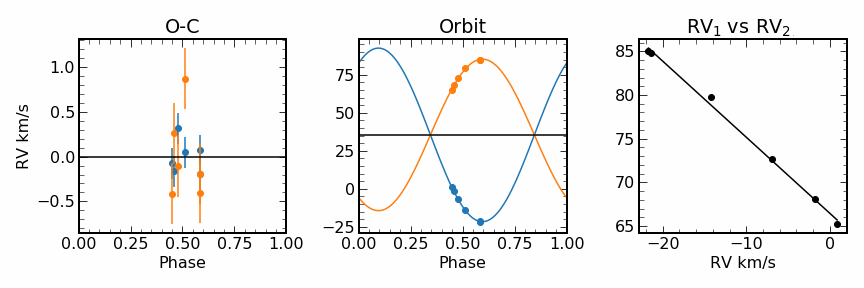}
    \caption{Comparison of the single-star (offset +0.5) and binary model fits for 18503840-0617048 (top panel). Purple line shows the error spectrum. Binary fit is done using mass ratio from orbital fitting. On bottom panels we show fit residuals O-C, phase folded RV curves and the Wilson plot.}
    \label{fig:fits0}
\end{figure*}

\begin{figure*}
	\includegraphics[width=\textwidth]{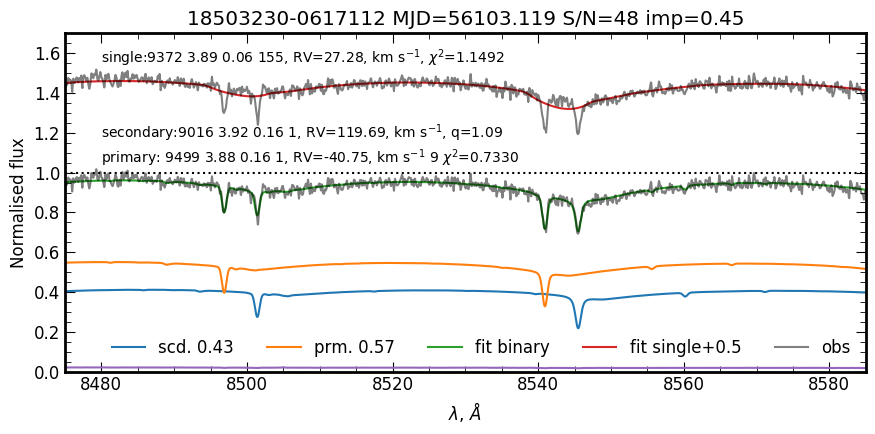}
	\includegraphics[width=\textwidth]{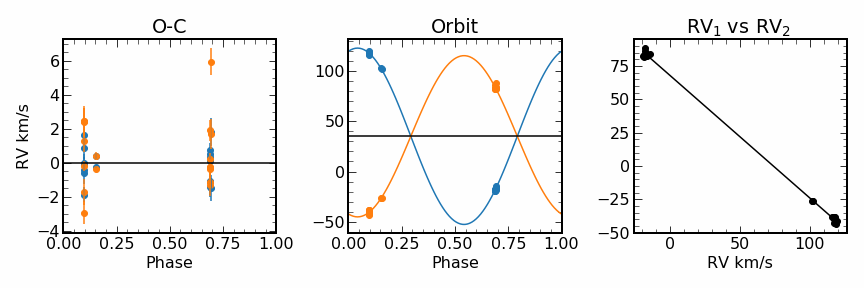}
    \caption{Same as Figure~\ref{fig:fits0} but for 18503230-0617112.}
    \label{fig:fits1}
\end{figure*}

\begin{figure*}
	\includegraphics[width=\textwidth]{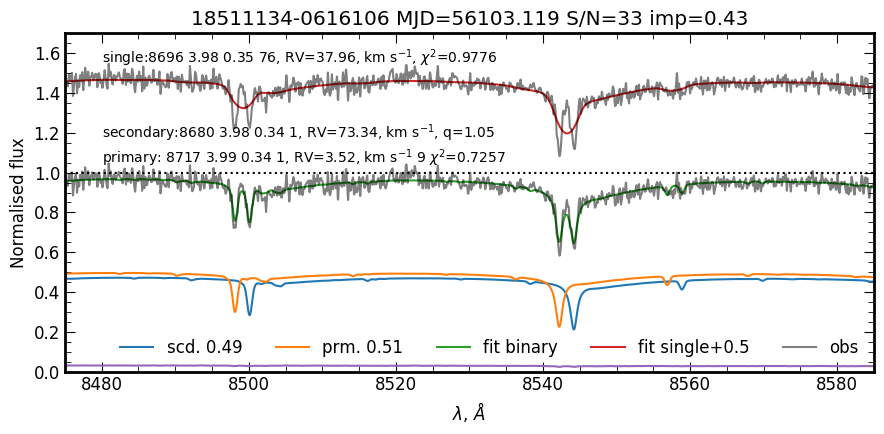}
	\includegraphics[width=\textwidth]{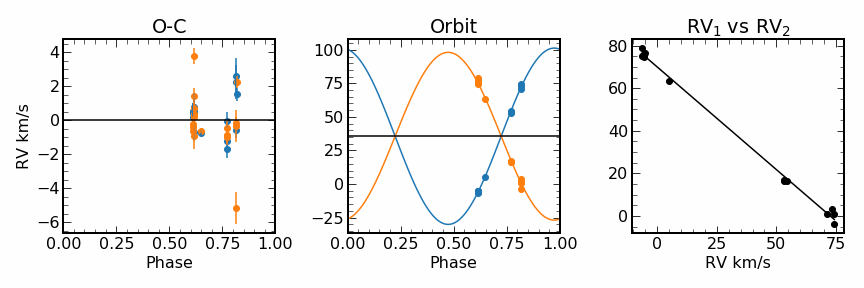}
    \caption{Same as Figure~\ref{fig:fits0} but for 18511134-0616106.}
    \label{fig:fits2}
\end{figure*}

\section{Results \& Discussion}
\label{results}
We have analysed all spectra using both binary and single star spectral models\footnote{Plots showing fits for all stars are available on \url{https://nlte.mpia.de/upload/kovalev/}}.
In the right panel of the Figure~\ref{fig:sel} we apply the selection from Section~\ref{classify} to the observed spectra. Only 9 stars have $\chi^2_{\rm single} < \chi^2_{\rm binary}$ and are not shown on this plot. Among the remaining $256$ stars we have $171$ possible binary candidates with $\log{\Delta \chi^2}>-2.5$. The six SB2 candidates from \citet{merle2017} are part of these $171$ stars. They are shown with blue circles. SB3 candidate from \citet{merle2017} (cyan star) is also selected. Among the 10 SB1 candidates from \citet{merle2020} (gray circles) only one is not selected. The remaining $85$ stars with $\log{\Delta \chi^2}<-2.5$ can still be binaries, potentially twins with very similar RVs.
\par
Compared with the test on synthetic spectra, we see a few stars with $\log{\Delta \chi^2}$ much larger than the general trend. Some such stars are known variables or have emission around the Ca~II lines and cannot be modelled properly. Similar chromospheric emission has been detected earlier in \citet{sb2rave}. Several other stars have binary solutions with extrapolations beyond the edges of the synthetic model grid. Obviously, the binary spectral model fits with additional five free parameters will have a smaller $\chi^2$ than the single-star model. Therefore, in order to make sure that our choice is correct, we carefully check all spectra by eye and choose them only if the spectra clearly show signs of the composite structure. We also add SB2s from \citet{merle2017} and \citet{kv29} and get a final set of $26$ SB2 candidates. We show them as yellow triangles in the Figure~\ref{fig:sel} and list their parameters in the Table~\ref{tab:my_res}. Almost all of these stars have $f_{\rm imp}>0.10$. In the Figure~\ref{fig:degenobs} we reproduce Figure~\ref{fig:degeracy} for observed spectra. All selected SB2s are shown with open triangles. Majority of the datapoints with $\Delta\rv$ correlated with $\vsini$ from the single-star fit are highly likely single stars, as the sum $\vsini_1+\vsini_2$ is quite large and $f_{\rm imp}<0.1$.
\par 
 The average metallicity for the selected SB2s is $\langle\feh\rangle=0.19\pm0.16$ dex, which is higher than the high-resolution spectroscopic estimate $\langle\feh\rangle=0.10\pm0.06$ dex by \citet{cantat2014}  and the photometric estimate $\langle\feh\rangle=0.04\pm0.06$ dex by \citet{recentgaia} for M~11 cluster. We also calculated $\langle\feh\rangle=0.23\pm0.15$ dex for 20 stars, which are presumed to be single-stars ($\log{\Delta \chi^2}<-2.5$) and cluster members ($\rv$ is within $3\sigma$ interval from $\langle\rv\rangle=35.9\pm2.8\kms$ \citep{cantat2014}). Thus, our $\feh$ estimates are biased relative to the literature values. We found no correlation of $\feh$ with $\vsini$ for these 20 stars, but observed that the dispersion is higher for hotter stars ($\feh=0.18\pm0.08$ dex for 6 stars with $\teff<7000$ K and $\feh=0.25\pm0.17$ dex for 14 stars with $\teff>7000$ K). We leave the detailed study of this bias for future studies, as it is beyond the scope of this paper.  
\par
Below we briefly describe a few of the individual groups on our SB2 list:
\begin{enumerate}
\item 18503840-0617048 is a pair of main-sequence stars with narrow and deep spectral lines,  see top panel of Figure~\ref{fig:fits0}. Not surprisingly, this system has the maximum improvement factor $f_{\rm imp}=0.49$.
\item 18512031-0609011, 18510456-0617121, 18503230-0617122  (see top panel of Figure~\ref{fig:fits1}), 18511134-0616106  (see top panel of Figure~\ref{fig:fits2}) and 18510223-0614547 are hot binaries and highly likely twin stars. The wide Ca~II lines are probably the reason why 18512031-0609011 was not included in the SB2 candidate list \citet{merle2017}, as the double lines of the spectrum are clearly visible. 
\item 18512155-0618391, 18511060-0619206, 18511459-0620280 and 18505270-0621406 represent systems with high rotational velocities of the  components. 
Stars 18505270-0621406 and 18511459-0620280 shows composite spectra which have very discrepant $\vsini$ for the primary and secondary components, see Figure~\ref{fig:fastrot}. 
 In these binaries one star rotates significantly faster $\Delta \vsini\sim 150,\,190~\kms$ than the other. A similar discrepancy has been observed in other stars, see \citet{2017AstBu..72...16Z,2018AstBu..73..351Z}.
\begin{figure*}
	\includegraphics[width=\textwidth]{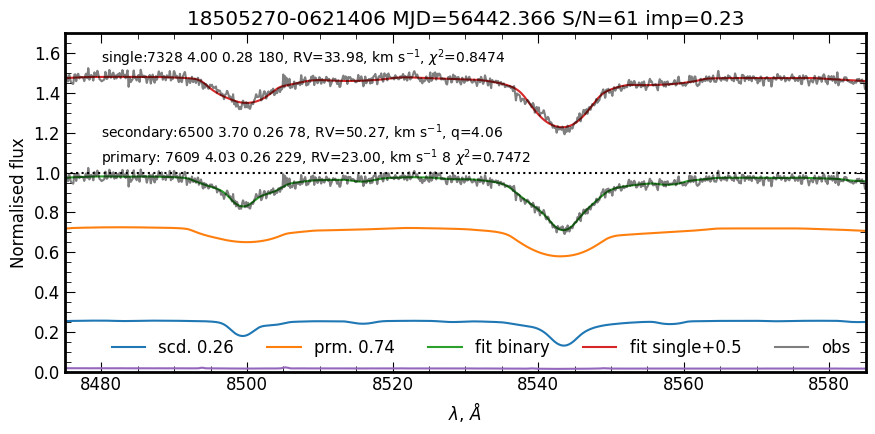}
	\includegraphics[width=\textwidth]{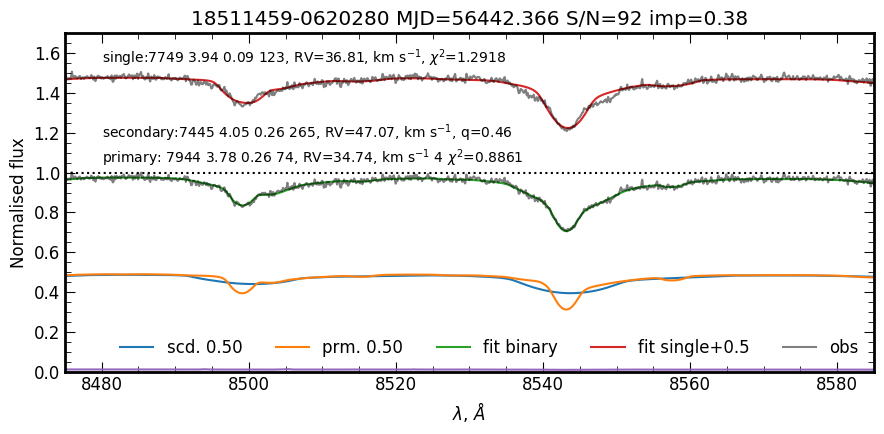}
    \caption{Fit of the spectrum for systems with discrepant $\vsini$ for the two components.}
    \label{fig:fastrot}
\end{figure*}
\item 18510579-0616398, 18502952-0614591, 18505557-0611345, 18505797-0617083, 18513592-0618453 and 18510492-0608570 are systems in which the secondary component is much dimmer (<15\%), but still clearly visible in the spectrum. 
Star 18505557-0611345 with $\Delta \vsini\sim 100\kms$ is an interesting case. We have a spectrum of a fast rotating primary star, with clearly visible narrow lines of the faint secondary component ($\loa 7\%$ of the light). 
We probably have a random alignment with the background/foreground star here, although our set of synthetic stars has systems with similar spectra. 
\item 18511836-0619458, 18514000-0619405, 18513636-0616190, 18512203-0609346, 18510656-0614562, 18510368-0617353, 18505693-0616214, 18505797-0617083, 18505557-0611345, 18505379-0613443 and 18502952-0614591 show no clear evidence of double structure, but have quite high improvement factor. 18502952-0614591 and 18505797-0617083 are listed as SB1 candidates in \citet{merle2020}. The radial velocities for both components in 18512203-0609346 are not consistent with its cluster membership.
\item poor results ($f_{\rm imp}<0.10$) for two previously known SB2s: 18510012-0616373 and  18510462-0616124, see description below.
\end{enumerate}

\subsection{Known SB2 candidates}

Our results confirm five of the six SB2 candidates discovered earlier in \citet{merle2017}. These stars show double lines in their spectra, and our binary model successfully derives radial velocities and spectral parameters (see top panels of Figures~\ref{fig:fits0},\ref{fig:fits1},\ref{fig:fits2}). Only star 18510462-0616124 shows no composite spectrum structure. It is likely that this spectrum was observed near the conjunction phase of the binaries. Nevertheless, we keep this star in the SB2 list as an example of a poor result.
\par
\begin{figure*}
	\includegraphics[width=\textwidth]{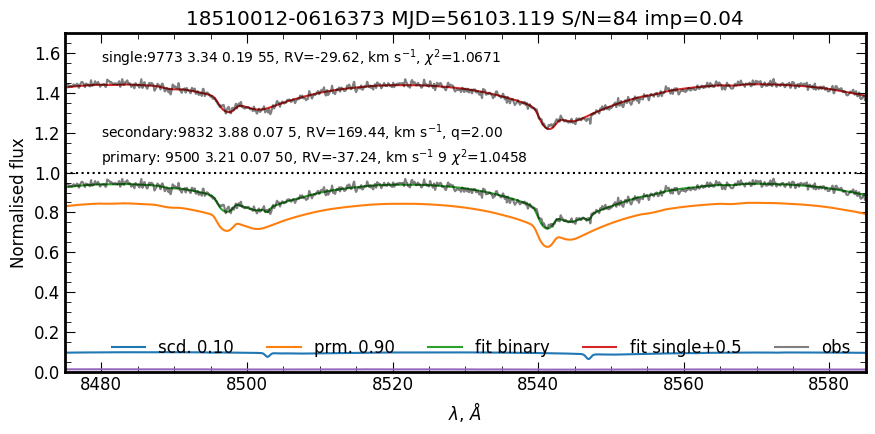}
    \caption{Comparison of single and binary model fits for the known eclipsing binary 18510012-0616373 (KV 29). The mass ratio from \protect\citet{kv29} is used in the binary model fitting.}
    \label{fig:kv29}
\end{figure*}

A detached eclipsing binary KV~29 is included in our sample with GES name 18510012-0616373. \citet{kv29} reports $\teff=9480\pm550$~K, $\logg=3.531\pm0.004$ cgs units, $\vsini=57.60\pm0.32\kms$ for the primary component and $\teff=7810\pm480$~K, $\logg=4.264\pm0.022$ cgs units, $\vsini=17.39\pm0.46\kms$ for the secondary component, mass ratio $Q=0.509\pm0.004$ ($q=1.96$) and circular orbit with period $P=4.64276\pm0.00001$ days. Our method is not able to get a proper proof that it is SB2, since the secondary component is very weak, so the binary solution recovers only the contribution of the primary component. So we repeat the analysis, using fixed mass ratio $q=2.00$ and find a significant improvement in the fit, although the solution is still not robust with $f_{\rm imp}<0.10$,  see Figure~\ref{fig:kv29} . The secondary component contributes only $\loa 10 \%$, but it is still visible in the spectrum. We find  $\teff=9500\pm27$~K, $\logg=3.21\pm0.01$ cgs units, $\vsini=50\pm1\kms$ for the primary component and  $\teff=9832\pm205$~K,$\logg=3.88\pm0.07$ cgs units, $\vsini=5\pm10\kms$ for the secondary component, which is different from \citet{kv29} especially for the secondary component. The orbital solution from \citet{kv29} $\rv_{1,2}= -38.10,\, 170.87\kms$ computed for HJD=2456103.625\footnote{Converted from MJD=56103.119 using Pyastronomy \citep{pyasl}} is close to our estimates $\rv_{1,2}=-37.24,\,169.4\kms$. 
 In any case, we keep this star on our SB2 list as an example of poor fit.

\subsection{SB3 candidate}
\begin{figure*}
	\includegraphics[width=2\columnwidth]{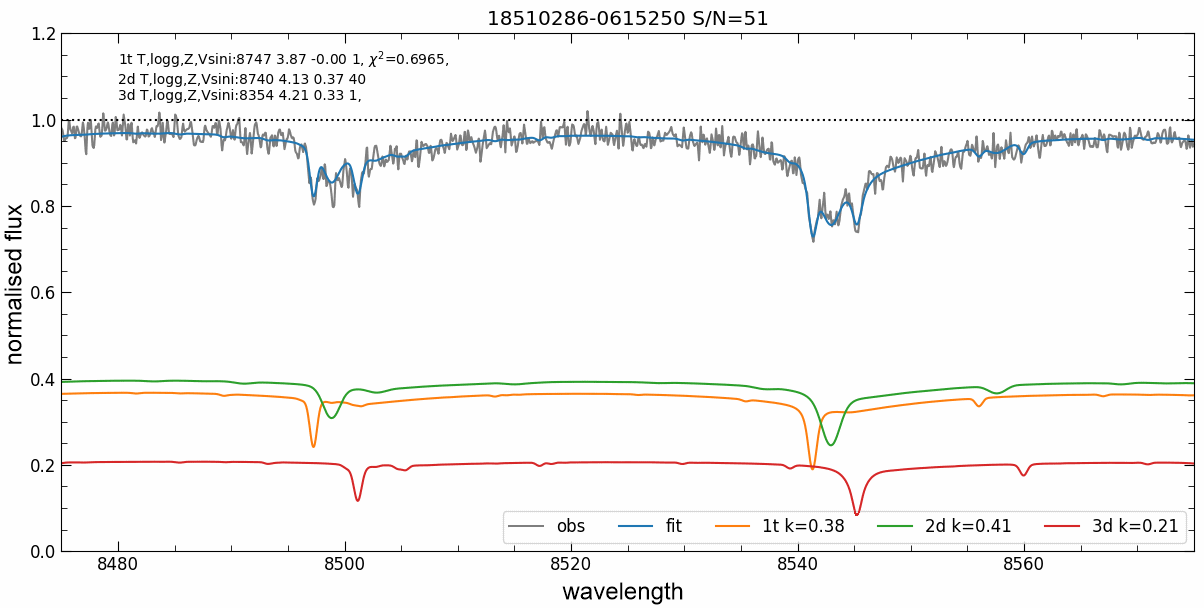}
    \caption{Fit for a star 18510286-0615250 with a triple spectrum.}
    \label{fig:tri}
\end{figure*}

Star 18510286-0615250 was identified in \citet{merle2017} as SB3 candidate based on a cross-correlation function peak analysis.
Upon visual inspection, we found that the spectrum of 18510286-0615250 is a composite of three spectra: a central spectrum with broad lines and two blue/redshifted components with narrow lines.  Our best fit to the binary model has extracted only the red-shifted component, while the two remaining components are fitted as one spectrum with very wide lines. So we treat this complex spectrum in a special way. We run the binary model fitting on the spectrum where the contribution from the redshifted component is subtracted. Thus, we have radial velocities for all spectral components: $\rv_{1,2,3}=-27.30,~29.31,~110.68 \kms $ (MJD=56103.119). Given fixed values of $\rv$s, we fit the original spectrum and obtain spectral parameter estimates for all three components, where all mass ratios are calculated with respect to the blue-shifted component. The final fit is shown in Figure~\ref{fig:tri}.

\par
The central component is the heaviest, contributing $41\%$ to the luminosity, the red-shifted component is lighter, contributing $21\%$ to the luminosity, and the blue-shifted component has the least mass, contributing $38\%$ to the luminosity. The central component rotates at $\vsini=40 \kms$, while the other two components show no sign of rapid rotation in their spectra. The blue-shifted component has solar metallicity, while the other two components are supersolar with $\feh \sim 0.35~$ dex. We calculated the center of mass radial velocity of the system using two mass ratios $\rv_{\rm bc}=42.36~\kms$, and it is only $\sim 7 \kms$ higher than the average radial velocity of the cluster. Thus, this triple system can be part of the cluster. Possible SB3 configuration is a hierarchical triple system of inner subsystem with mass ratio $q_{\rm inner}=\frac{m_3}{m_1}=1.39$ and outer subsystem with mass ratio $q_{\rm outer}=\frac{m_1+m_3}{m_2}=1.23$.
\par
\citet{merle2017} also studied 18510286-0615250 as an SB3 candidate. However, their analysis of spectra of several epochs showed that the central component is not gravitationally bound to the other two, since its $\rv$ does not change much with time. Taking this into account, we calculated the center of mass radial velocity without the central component and obtained $\rv_{\rm bc}=52.95~\kms$, which is $\sim 18 \kms$ more than the average radial velocity of the cluster. However, our mass ratios are not very reliable to draw rigorous conclusions, so the question about the status of SB3 is still open.

\subsection{Orbital solutions}
We expect that for double-lined binaries we can get orbital solutions, so we select three SB2 candidates with high $f_{\rm imp}>0.4$ and extract for them all available GIRAFFE spectra from the ESO archive\footnote{\url{https://archive.eso.org}}. These spectra were collected using different wavelength settings on five nights during four years (MJD=56077, 56099, 56103, 56442, 57536). We compute six high resolution synthetic spectral templates in the wavelength range $\lambda=4000, 6800$~\AA~with the best-fit spectral parameters from the binary results using the NLTE MPIA\footnote{\url{https://nlte.mpia.de}} spectral synthesis interface \citep[see Chapter 4 in][]{disser}. We discard all spectra taken on MJD=56077 as very noisy, so we only have a total of 4 nights with a few observations, and thus phase coverage will be insufficient. We fit the radial velocities for both components of the binary to appropriate templates in the same way as we did for the HR21 spectra, except here we keep the spectral parameters fixed and fit the radial velocities and resolution of the observed spectrum assuming a Gaussian instrumental profile. We find only 5, 17 and 16 reliable RV estimates for the systems 18503840-0617048, 18503230-0617112 and 18511134-0616106 respectively. However, it should be noted that these radial velocities are less accurate than those obtained by the full binary fitting described in Section~\ref{sec:maths}. 
\par
If binary components have similar spectra, their RVs can be flipped up, so we reassign components for RV estimations, requiring that all RVs are similar if they were obtained on the same night. We check this assignment by plotting the RVs of each component relative to each other on the Wilson plot \citep{wilson}. If the two stars are gravitationally bound all data points will follow a straight line (see Formula~\ref{eq:asgn}). The correct assignment is very important in orbital fitting.  
\par
In the next step radial velocities are used to fit circular orbits using following formula:
\begin{align}
    \label{eq:orbit}
    {\rm RV_2}(t)=\gamma- K_2 \sin \left (\frac{2\pi}{P}(t -t_0) \right ),\\
    {\rm RV_1}(t)=\gamma+ K_1 \sin \left (\frac{2\pi}{P}(t -t_0) \right ),\\
    q_{\rm dyn}=\frac{m_1}{m_2}=\frac{K_2}{K_1},
\end{align}
where $\gamma$ is the systemic velocity, $P$ is the period, $t_0$ is the time of conjunction, $K_n$ are radial velocity amplitudes for $n$-component. In order to explore the parameter space and avoid local minima, we perform the optimisation several times with different optimiser initialisations. As a final result, we choose the solution with the minimal $\chi^2$. Kepler's third law and inclination of the orbit allow us to constrain the total mass $m_{\rm tot}$ and radius of the orbit $a=(a_1+a_2)$, using the period and amplitudes of the radial velocity:

\begin{align}
    (m_1+m_2)\sin^3{i}=\frac{(K_1+K_2)^3}{GM_\odot} \frac {P}{2 \pi} ,\\
    (a_1+a_2)\sin{i}=(K_1 + K_2)\frac{P}{2 \pi} ,
\end{align}
where $GM_\odot=1.32712440041\cdot 10^{20}\, {\rm m^3\,s^{-2}}$  is the Solar mass parameter\footnote{\url{https://iau-a3.gitlab.io/NSFA/NSFA_cbe.html\#GMS2012}}, $i$ is the inclination of the orbit to the sky-plane. The masses of the components can be found using the total mass and $q_{\rm dyn}$. 
\par
 Results for orbital fitting are shown in bottom panels of Figures~\ref{fig:fits0},\ref{fig:fits1},\ref{fig:fits2}. We have collected all fitted and derived quantities in Table~\ref{tab:orbits}. All three systems are close binaries with periods $\sim 1-3$ weeks and their systemic velocities are consistent with the average cluster velocity.  Assuming the highest orbital inclination $i=90^{\circ}$ the system 18503840-0617048 consists of sun-like stars that are $\sim~30\,R_\odot$ from each other. Systems 18503230-0617112 and 18511134-0616106 have components nearly twice as heavy as the Sun and are located $\sim~30,\,50\,R_\odot$ apart.  
\par 
Dynamic mass ratios $q_{\rm dyn}$ are more reliable than spectroscopic ratios $q$ (based on the results of a test over a mock spectra), so we use them to re-fit the spectra of binaries with a fixed $q=q_{\rm dyn}$. For 18503840-0617048 this decreased the $\logg$ of the cold component by 0.5 dex, while for the other two systems it made the secondary component parameters similar to the primary.  
\par
It should be noted that our analysis of orbits is not very reliable, as the phase coverage for all three stars is poor and the circular orbit may be a poor approximation.  Also it should be noted that there is a possible aliasing on the periods due to few number of epochs, that makes the errors on the period $P$ of Table~\ref{tab:orbits} certainly underestimated. Additional observations for these stars will be very useful for further analysis.

\begin{table*}
    \centering
    \caption{The results of the circular orbit fitting.}
    \begin{tabular}{ccccccccc}
\hline
\hline
star &$P$, &$t_0$, &$K_1$, &$K_2$, &$\gamma$, &$a\sin i,$ &$m_{\rm tot}\sin^3 i, $&$\frac{m_1}{m_2}$ \\
N spectra & err, d& err, d& err, $\kms$& err, $\kms$& err, $\kms$& err, $R_\odot$& err, $M_\odot$&err\\
\hline
\hline
18503840-0617048& 14.7967 & 56417.649 & 49.86 & 57.14 & 35.48 & 31.3 &  1.88 & 1.15\\
6 & 0.0007 & 1.126 & 0.59 & 0.48 & 0.43 & 0.3 & 0.06 & 0.02\\
\hline
18503230-0617112& 9.4242 & 56072.449 & 80.19 & 87.71 & 35.06 & 31.3 & 4.62 & 1.09\\
18 & 0.0001 & 0.029 & 0.37 & 0.36 & 0.07 & 0.1 & 0.06 & 0.01\\
\hline
18511134-0616106& 18.8975 & 56091.730 & 62.51 & 65.68 & 35.61 & 47.9 &  4.12 & 1.05\\
17 & 0.0001 & 0.104 & 0.20 & 0.21 & 0.08 & 0.1 & 0.04 & 0.01\\
\hline
    \end{tabular}
    \label{tab:orbits}
\end{table*}


\subsection{Comparison with Gaia data.}
\label{discus}
\begin{figure}
	\includegraphics[width=\columnwidth]{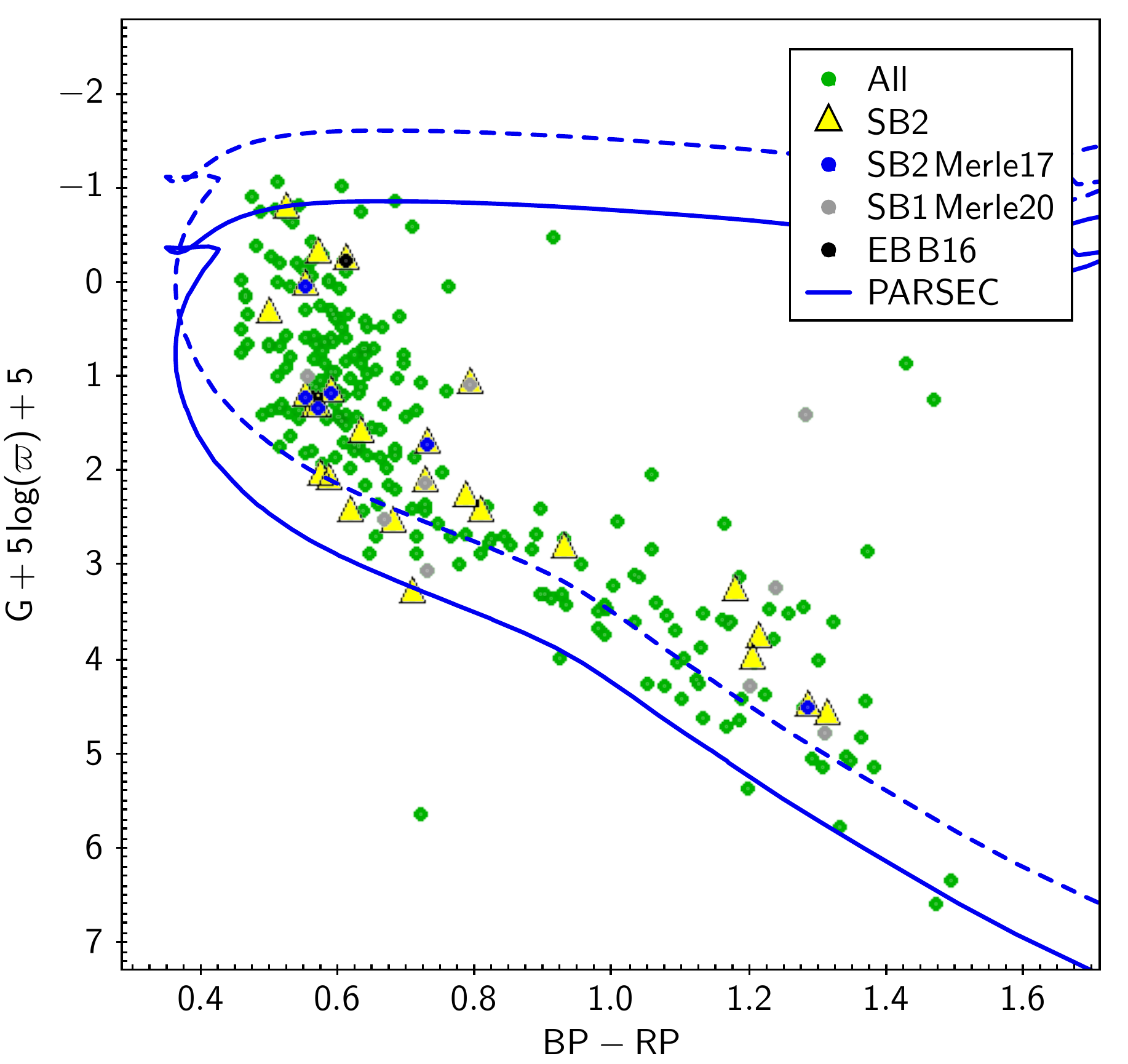}
    \caption{Hertzsprung-Russell diagram from Gaia eDR3 data. The colours and symbols are the same as in the Figure~\ref{fig:sel}. The PARSEC isochrone is shown as a solid blue line. We shift it according to the reddening  $E(B-V)=0.47$ mag and the extinction $A_V=0.94$ mag. The same isochrone for twin binary stars is shown $0.75$ mag above by the dashed blue line.}
    \label{fig:hrd}
\end{figure} 
Figure~\ref{fig:hrd} shows the Hertzsprung-Russell diagram calculated from Gaia eDR3 data \citep[][]{egdr3} for 260 stars with positive parallax. SB3 candidate has negative parallax and is not shown. We use the same colours and symbols as in Figure~\ref{fig:sel}. The PARSEC isochrone used to generate mock stars in Section~\ref{classify} is shown as a solid blue line. We shift it according to the reddening $E(B-V)=0.47$ mag (WEBDA database\footnote{\url{https://webda.physics.muni.cz/}}) and the extinction $A_V=2E(B-V)=0.94$ mag. The same isochrone for twin binary stars is shown $0.75$ mag above as a dashed blue line. We can see that many of the detected SB2s candidates are higher than main sequence single stars. SB2s are brighter than single stars, so Gaia eDR3 data qualitatively confirm our results.

\section{Conclusions}
\label{concl}
We have developed a new method for the analysis of binary spectra.
Our method successfully identifies SB2 candidates from high-resolution Gaia-ESO spectra. Compared to the commonly used CCF analysis, it also works for binaries with rapidly rotating components. We use spectral fitting but, unlike \cite{bardy2018}, we do not take information from stellar isochrones to characterise binary components. Unfortunately, mass ratios obtained spectroscopically are not reliable. However, the use of multiple epoch spectra can solve this problem (see Kovalev et al. in prep). We have tested our method on synthetic and observed spectra of BAFG-stars in the open cluster M~11. We confirm five previously detected SB2 candidates and find 19 new ones.  For three SB2 candidates we fit circular orbits and obtain dynamical mass ratios. These mass ratios allow us to correct previous estimates of spectral parameters. We hope that our method will be useful in future analyses of high and intermediate resolution spectra, e.g. those obtained by Gaia RVS \citep{grvs}.

\section*{Data Availability}
The data underlying this article will be shared on reasonable request to the corresponding author.

\section*{Acknowledgements}
We are grateful to the anonymous referee for a constructive report. Mikhail Kovalev is grateful to his parents, Yuri Kovalev and Yulia Kovaleva, for their full support in making this research possible. We thank Hans B{\"a}hr for his careful proof-reading of the manuscript.
The work is supported by the Natural Science Foundation of China (Nos. 11733008, 12090040, 12090043).
 The research leading to these results has (partially) received funding from the European Research Council (ERC) under the European Union’s Horizon 2020 research and innovation programme (grant agreement \textnumero 670519: MAMSIE) and from the KU Leuven Research Council (grant C16/18/005: PARADISE).
Based on data products from observations made with ESO Telescopes at the La Silla Paranal Observatory under run IDs 188.B-3002 and 193.B-0936.
This research made use of the SIMBAD database, operated at CDS, Strasbourg, France.
This work has made use of the VALD database, operated at Uppsala University, the Institute of Astronomy RAS in Moscow, and the University of Vienna. 
This research has made use of the WEBDA database, operated at the Department of Theoretical Physics and Astrophysics of the Masaryk University. 
It also made use of TOPCAT, an interactive graphical viewer and editor for tabular data \citep[][]{topcat}.




\bibliographystyle{mnras}
 


\appendix

\section{List of detected SB2 candidates}

\begin{table*}
    \centering
    \begin{tabular}{c|cccc|cccc|cccc}
\hline
\hline
star &  \multicolumn{4}{c}{primary} & \multicolumn{4}{c}{secondary}&\multicolumn{4}{c}{}\\
cname & RV & $\teff$ & $\logg$ & $\vsini$& RV & $\teff$ & $\logg$ & $\vsini$& $\feh$ & q & $f_{\rm imp}$ & frac \\
 info&$\kms$& K & cgs & $\kms$&$\kms$& K & cgs & $\kms$ & dex & low..high & $\Delta \chi^2$& S/N\\
\hline 
\hline
18502952-0614591 & 33.93 & 7991 & 3.39 & 1 & -39.82& 7709 & 4.49 & 36 & 0.18 & 1.12 & 0.18 &0.94 \\
M20  & (0.08) &(11) &(0.01) &(1) & (2.65) &(217) &(0.21) &(6) &(0.01) &0.70..1.80 & 0.05& 67 \\
18503230-0617112 & -40.75 & 9499 & 3.88 & 1 & 119.69& 9016 & 3.92 & 1 & 0.16 & 1.09$^{\rm OF}$ & 0.45 &0.57 \\
M17  clear CS OF & (0.30) &(43) &(0.03) &(3) & (0.34) &(80) &(0.04) &(3) &(0.03) &-- & 0.42& 48 \\
18503840-0617048 & 79.80 & 6481 & 4.44 & 1 & -14.25& 5649 & 4.59 & 1 & 0.19 & 1.15$^{\rm OF}$ & 0.49 &0.71 \\
M17  clear CS OF & (0.17) &(40) &(0.05) &(2) & (0.34) &(77) &(0.16) &(3) &(0.01) &-- & 0.68& 32 \\
18505270-0621406 & 23.00 & 7609 & 4.03 & 229 & 50.27& 6500 & 3.70 & 78 & 0.26 & 4.06 & 0.23 &0.74 \\
 clear CS & (1.35) &(37) &(0.05) &(3) & (1.07) &(91) &(0.16) &(3) &(0.01) &2.61..6.31 & 0.10& 61 \\
18505379-0613443 & 37.26 & 8773 & 3.87 & 235 & 29.17& 6000 & 3.74 & 21 & 0.37 & 5.70 & 0.12 &0.92 \\
 & (1.60) &(56) &(0.02) &(6) & (1.78) &(308) &(0.64) &(5) &(0.05) &1.36..23.86 & 0.02& 34 \\
18505557-0611345 & 37.01 & 8666 & 3.96 & 100 & 66.43& 6097 & 3.35 & 1 & 0.29 & 23.37 & 0.14 &0.93 \\
 clear CS & (0.59) &(25) &(0.01) &(1) & (0.75) &(140) &(0.29) &(7) &(0.01) &11.87..46.02 & 0.04& 70 \\
18505693-0616214 & 11.30 & 9938 & 3.80 & 40 & 71.45& 10000 & 3.72 & 78 & 0.24 & 1.46 & 0.19 &0.55 \\
 & (0.92) &(60) &(0.04) &(2) & (9.66) &(94) &(0.05) &(12) &(0.03) &1.01..2.12 & 0.04& 66 \\
18505797-0617083 & 34.45 & 9392 & 4.12 & 29 & 28.81& 6463 & 3.40 & 93 & 0.52 & 14.05 & 0.29 &0.87 \\
M20  & (0.16) &(19) &(0.01) &(<1) & (1.36) &(87) &(0.15) &(3) &(0.01) &9.90..19.94 & 0.19& 94 \\
18510012-0616373 & -37.24 & 9500 & 3.21 & 50 & 169.44& 9832 & 3.88 & 5 & 0.07 & 2.00$^{\rm B16}$ & 0.04 &0.90 \\
B16 poor result & (0.41) &(27) &(0.01) &(1) & (1.16) &(205) &(0.07) &(10) &(0.02) &-- & 0.02& 84 \\
18510223-0614547 & 43.74 & 9735 & 3.89 & 1 & 13.88& 9422 & 4.00 & 4 & 0.45 & 2.61 & 0.25 &0.78 \\
M17  clear CS & (0.35) &(57) &(0.03) &(2) & (1.16) &(115) &(0.12) &(9) &(0.02) &1.80..3.78 & 0.05& 53 \\
18510368-0617353 & 48.90 & 10500 & 3.75 & 28 & 4.46& 8484 & 3.93 & 22 & 0.32 & 4.12 & 0.14 &0.91 \\
 & (0.55) &(45) &(0.01) &(1) & (2.07) &(154) &(0.20) &(6) &(0.02) &2.55..6.64 & 0.03& 65 \\
18510456-0617121 & -9.47 & 9869 & 3.75 & 1 & 81.94& 9982 & 3.79 & 1 & 0.04 & 1.07 & 0.41 &0.54 \\
M17  clear CS & (0.39) &(97) &(0.05) &(3) & (0.47) &(115) &(0.05) &(4) &(0.04) &0.84..1.37 & 0.17& 48 \\
18510462-0616124 & 41.96 & 10071 & 3.60 & 16 & 54.51& 9992 & 3.68 & 167 & 0.36 & 2.82 & 0.04 &0.78 \\
M17 poor result & (0.21) &(40) &(0.03) &(1) & (2.93) &(197) &(0.12) &(13) &(0.02) &2.03..3.92 & 0.02& 77 \\
18510492-0608570 & -10.22 & 9373 & 3.94 & 29 & 49.02& 6482 & 4.27 & 9 & -0.07 & 1.14 & 0.19 &0.86 \\
 clear CS & (0.55) &(44) &(0.01) &(1) & (0.87) &(153) &(0.28) &(7) &(0.04) &0.61..2.13 & 0.08& 40 \\
18510579-0616398 & 7.28 & 9998 & 3.39 & 1 & 100.91& 8993 & 3.55 & 11 & 0.52 & 4.46 & 0.15 &0.89 \\
 clear CS & (0.09) &(17) &(0.01) &(1) & (0.61) &(80) &(0.06) &(2) &(0.01) &3.80..5.24 & 0.19& 101 \\
18510656-0614562 & 29.03 & 10002 & 3.32 & 28 & 40.77& 9980 & 3.59 & 133 & 0.34 & 0.87 & 0.14 &0.61 \\
 & (0.29) &(59) &(0.04) &(1) & (1.11) &(108) &(0.07) &(4) &(0.02) &0.68..1.11 & 0.08& 113 \\
18511060-0619206 & 47.65 & 7888 & 4.07 & 117 & 17.75& 6445 & 4.14 & 36 & 0.19 & 1.89 & 0.25 &0.79 \\
 clear CS & (0.88) &(22) &(0.02) &(1) & (0.68) &(73) &(0.14) &(2) &(0.01) &1.35..2.64 & 0.11& 68 \\
18511134-0616106 & 3.52 & 8717 & 3.99 & 1 & 73.34& 8680 & 3.98 & 1 & 0.34 & 1.05$^{\rm OF}$ & 0.43 &0.51 \\
M17  clear CS OF & (0.35) &(66) &(0.05) &(3) & (0.36) &(67) &(0.06) &(4) &(0.03) &-- & 0.25& 33 \\
18511459-0620280 & 34.74 & 7944 & 3.78 & 74 & 47.07& 7445 & 4.05 & 265 & 0.26 & 0.46 & 0.38 &0.50 \\
 clear CS & (0.47) &(47) &(0.04) &(1) & (1.40) &(62) &(0.06) &(3) &(0.01) &0.37..0.58 & 0.41& 92 \\
18511836-0619458 & 40.20 & 6868 & 4.35 & 89 & 27.52& 6005 & 4.77 & 10 & 0.14 & 1.70 & 0.14 &0.87 \\
 & (0.51) &(30) &(0.03) &(1) & (0.68) &(179) &(0.33) &(6) &(0.01) &0.81..3.57 & 0.03& 49 \\
18512031-0609011 & 87.85 & 9377 & 4.14 & 28 & -76.03& 9244 & 3.92 & 36 & -0.20 & 1.83 & 0.38 &0.53 \\
 clear CS & (0.55) &(55) &(0.02) &(1) & (0.76) &(63) &(0.03) &(2) &(0.03) &1.59..2.11 & 0.22& 65 \\
18512155-0618391 & 21.48 & 8861 & 4.04 & 96 & 57.42& 7000 & 3.85 & 29 & 0.24 & 2.72 & 0.32 &0.76 \\
 clear CS & (0.74) &(25) &(0.01) &(1) & (0.42) &(38) &(0.07) &(1) &(0.01) &2.25..3.29 & 0.28& 92 \\
18512203-0609346 & 24.77 & 7223 & 5.00 & 101 & -0.36& 5961 & 5.00 & 28 & 0.16 & 1.18 & 0.19 &0.67 \\
 clear CS & (1.26) &(51) &(0.07) &(2) & (0.63) &(71) &(0.17) &(1) &(0.02) &0.71..1.97 & 0.07& 39 \\
18513592-0618453 & -31.17 & 9927 & 3.95 & 48 & 166.92& 6565 & 4.37 & 11 & 0.47 & 1.96 & 0.17 &0.93 \\
 clear CS & (0.69) &(50) &(0.01) &(1) & (1.31) &(264) &(0.43) &(5) &(0.04) &0.77..5.05 & 0.07& 43 \\
18513636-0616190 & 64.97 & 6934 & 4.41 & 48 & 14.81& 5830 & 4.60 & 23 & 0.13 & 0.67 & 0.17 &0.63 \\
 & (1.57) &(52) &(0.06) &(2) & (0.93) &(68) &(0.12) &(1) &(0.02) &0.49..0.91 & 0.04& 35 \\
18514000-0610405 & 35.93 & 8648 & 3.44 & 25 & 11.55& 8986 & 4.33 & 203 & 0.61 & 0.21 & 0.17 &0.60 \\
 & (0.23) &(47) &(0.02) &(1) & (2.40) &(54) &(0.03) &(5) &(0.01) &0.19..0.23 & 0.16& 65 \\
\hline
    \end{tabular}
    \caption{List of all SB2 candidates. cname is encoded sky position ($\alpha,\delta$) HHMMSSss-DDMMSSs of the star, values in parentheses are standard errors, for spectroscopic mass ratios $q$ we provide interval $q_{\rm low}..q_{\rm high}$, frac=$\frac{k}{k+1}$ is the light contribution of the primary component. B16 is eclipsing binary with fit initialised using mass ratio from \protect\citet{kv29}, M17 is a SB2 candidate in \protect\citet{merle2017}, M20 is a SB1 candidate in \protect\citet{merle2020}, CS - composite spectrum, OF indicates stars with mass ratio from orbital fits.}
    \label{tab:my_res}
\end{table*}


\bsp	
\label{lastpage}
\end{document}